# Why Learners Drift In and Out: Examining Intermittent Discontinuance in AI-Mediated Informal Digital English Learning (AI-IDLE) Using SEM and fsQCA

**First Author and Corresponding Author**
Yiran Du
University of Cambridge, Cambridge, UK
yd392@cam.ac.uk

**Second Author**
Huimin He
Xi'an Jiaotong-Liverpool University, Suzhou, China
Huimin.he@xjtlu.edu.cn

**Abstract**
This study examined intermittent discontinuance in AI-mediated informal digital learning of English (AI-IDLE) through the cognition–affect–conation framework. Survey data were collected from 632 Chinese university EFL learners with prior AI-IDLE experience and analysed using structural equation modelling and fuzzy-set qualitative comparative analysis. The SEM results showed that perceived intelligence, perceived interactivity, and perceived personalisation reduced AI-IDLE intermittent discontinuance indirectly through enjoyment, whereas perceived ineffectiveness, perceived uncontrollability, and perceived complexity increased discontinuance indirectly through boredom. The fsQCA results further identified four configurational pathways leading to intermittent discontinuance, indicating that learners' temporary withdrawal from AI-IDLE can result from different combinations of cognitive barriers and affective disengagement. These findings extend AI-IDLE research from adoption and continuance to post-adoption discontinuance and highlight the need to design AI-supported English learning experiences that are enjoyable, personalised, controllable, and cognitively manageable.

## 1. Introduction

Informal digital learning of English (IDLE) refers to learners' self-directed English learning through digital technologies beyond formal instruction (Y. Zhang & Liu, 2025). With recent advances in artificial intelligence, AI-mediated informal digital learning of English (AI-IDLE) has emerged, enabling learners to use tools such as ChatGPT, automated feedback systems, speech-recognition applications, and adaptive platforms for autonomous English learning (Zadorozhnyy & Lee, 2025). These AI tools may provide intelligent, interactive, personalised, and immediate support, thereby enhancing learners' motivation, engagement, and English learning experiences (Liu, Darvin, & Ma, 2024; Yang et al., 2025).

However, AI-IDLE engagement is not always stable. Because it usually occurs in voluntary and self-regulated contexts, learners may temporarily stop using AI tools and later return to them. This behaviour, known as intermittent discontinuance, reflects cyclical movement between use, non-use, and reuse. It may occur when learners perceive AI tools as ineffective, uncontrollable, complex, repetitive, or poorly aligned with their learning needs (Liu, Darvin, & Ma, 2025). Although prior AI-IDLE research has examined adoption, motivation, enjoyment, and learning outcomes (Lee et al., 2024; Liu & Ma, 2024; Xia & Guo, 2025), intermittent discontinuance remains underexplored.

To address this gap, this study examines AI-IDLE intermittent discontinuance through the cognition–affect–conation framework, which explains behaviour as a process linking cognitive evaluations, affective responses, and behavioural tendencies (Zhou & Zhang, 2024). Specifically, it investigates how positive perceptions of AI tools reduce discontinuance through enjoyment, and how negative perceptions increase discontinuance through boredom. Using survey data from Chinese university EFL learners, this study combines structural equation modelling and fuzzy-set qualitative comparative analysis to identify both net effects and configurational mechanisms of AI-IDLE intermittent discontinuance (Kline, 2023; Schneider & Wagemann, 2012; Zhou & Zhang, 2025).

## 2. Literature Review

### 2.1 AI-Mediated Informal Digital English Learning (AI-IDLE)

Informal digital learning of English (IDLE) has been conceptualised as learners' self-directed and out-of-class engagement with digital technologies and online resources for English learning purposes, including activities such as watching videos, interacting on social media, gaming, using online communities, and accessing multimodal English input beyond formal instruction (Y. Zhang & Liu, 2025). Recent developments in artificial intelligence have extended this line of inquiry by shifting attention from general digital exposure to AI-mediated informal digital learning of English (AI-IDLE), in which learners draw on AI systems such as ChatGPT, Bing Chat, automated feedback tools, speech-recognition applications, and adaptive platforms to support autonomous English learning (Zadorozhnyy & Lee, 2025). Compared with earlier IDLE environments, AI-IDLE is distinguished by the perceived intelligence, interactivity, personalisation, immediacy, and conversational responsiveness of AI systems, which may provide learners with instant feedback, adaptive scaffolding, error correction, language practice, and affectively engaging interaction (Yang et al., 2025).

Emerging research suggests that AI-IDLE can strengthen learners' motivation, vocabulary development, engagement, and willingness to use English outside the classroom; however, the literature has largely focused on adoption, acceptance, and learning benefits, while paying less attention to unstable or interrupted patterns of use (Xia & Guo, 2025). This limitation is important because AI-IDLE occurs largely in informal, self-regulated contexts where learners are not institutionally required to continue using AI tools (Liu, Ma, et al., 2025). Therefore, understanding AI-IDLE requires not only examining why learners adopt or benefit from AI tools, but also why they temporarily withdraw from them despite having previously used them.

### 2.2 Intermittent Discontinuance

Intermittent discontinuance has been examined in information systems research as a post-adoption behaviour in which users temporarily suspend the use of a technology and later resume or readopt it, thereby differing from permanent discontinuance, replacement discontinuance, or complete rejection. Existing studies have shown that intermittent discontinuance is particularly relevant in voluntary digital environments, where use is shaped by changing perceptions of usefulness, emotional fatigue, cognitive burden, dissatisfaction, overload, and contextual needs. Unlike continuance research, which assumes relatively stable sustained use after adoption, discontinuance research highlights the fragility of post-adoption engagement and the possibility that users may cycle between use, non-use, and reuse. In educational technology contexts, this distinction is especially salient because learners' engagement with digital tools may fluctuate according to learning goals, perceived progress, emotional experience, platform complexity, and perceived control.

Within AI-IDLE, intermittent discontinuance can occur when learners perceive AI tools as ineffective, repetitive, uncontrollable, cognitively demanding, or insufficiently aligned with their English learning needs, even if they previously found such tools useful or enjoyable (Liu, Darvin, et al., 2025). Conversely, learners may return to AI-IDLE when new communicative needs arise, when the system provides more personalised support, or when positive affective experiences are restored (Liu, Zou, et al., 2025). However, prior AI-IDLE studies have rarely treated discontinuance as a focal outcome, leaving limited understanding of how cognitive and affective factors jointly explain learners' temporary withdrawal from AI-mediated informal English learning (Liu & Soyoof, 2026). Addressing intermittent discontinuance therefore broadens AI-IDLE research from adoption and continuance to a more dynamic account of learner–AI engagement over time.

## 3. Theoretical Framework and Hypothesis Development

### 3.1 Cognition–Affect–Conation Framework

The cognition–affect–conation framework provides a suitable theoretical basis for this study because it explains behaviour as a sequential process in which individuals' beliefs shape their emotional responses, which in turn influence behavioural tendencies (Zhou & Zhang, 2024). This logic is particularly relevant to AI-mediated informal digital learning of English (AI-IDLE), where learners' continued or interrupted use of AI tools depends not only on their cognitive evaluations of the technology, but also on the affective experiences generated during use (Liu & Ma, 2024). In this study, cognitive factors include perceived intelligence, perceived interactivity, perceived personalisation, perceived ineffectiveness, perceived uncontrollability, and perceived complexity; affective factors include enjoyment and boredom; and the conative outcome is intermittent discontinuance. Accordingly, the conceptual model presented in Figure 1 specifies how these cognitive and affective constructs are associated with learners' intermittent discontinuance in AI-IDLE.

**Figure 1. The Conceptual Model**

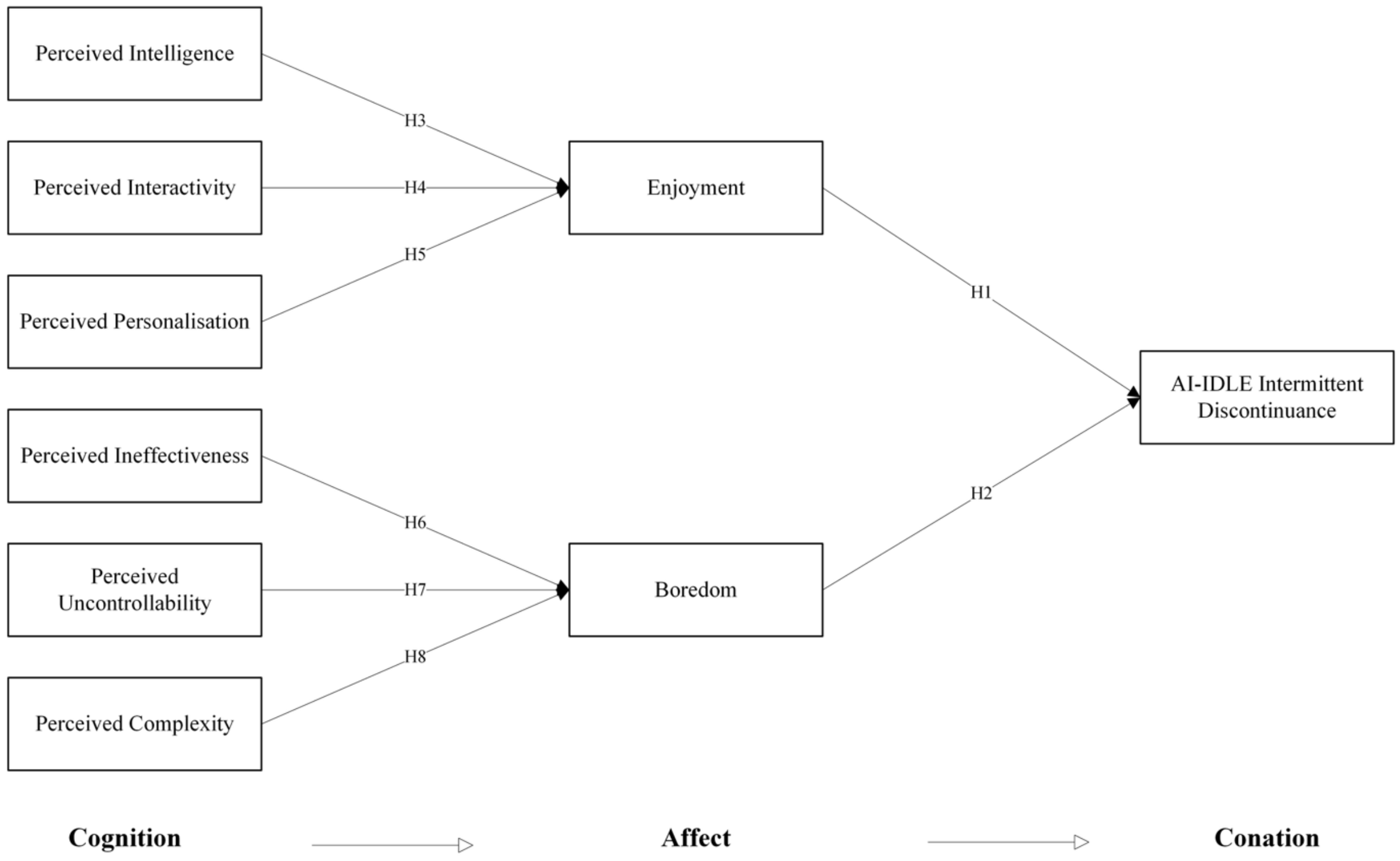

### 3.2 Enjoyment and Boredom

Enjoyment and boredom are included in the conceptual model because they represent two central affective experiences that may explain learners' voluntary engagement or withdrawal in AI-mediated informal digital learning of English (AI-IDLE) (Liu, Zou, et al., 2025). As AI-IDLE typically occurs outside formal instruction and is largely self-regulated, learners' emotional responses may play an important role in determining whether they continue using AI tools or temporarily suspend use (Lee et al., 2024). Enjoyment reflects pleasure, interest, and intrinsic satisfaction during AI-IDLE, which can make AI-mediated English learning feel rewarding and encourage repeated engagement (Liu, Zou, et al., 2025). When learners experience AI tools as interactive, responsive, and useful for their learning needs, enjoyment may reduce the likelihood of AI-IDLE intermittent discontinuance (W. Wu & Kabilan, 2025).

Boredom, by contrast, reflects low stimulation, reduced interest, and a perceived lack of meaningful engagement (Li et al., 2023). In AI-IDLE, boredom may occur when interactions with AI tools become repetitive, predictable, insufficiently challenging, or poorly aligned with learners' English learning goals. Because learners' use of AI tools in informal settings is voluntary, boredom can weaken their motivation to continue and increase the likelihood of temporary withdrawal (J. Zhang & Li, 2025). Therefore, boredom is expected to increase AI-IDLE intermittent discontinuance. Accordingly, the following hypotheses are proposed:

H1: Enjoyment is negatively associated with AI-IDLE intermittent discontinuance.
H2: Boredom is positively associated with AI-IDLE intermittent discontinuance.

### 3.3 Perceived Intelligence, Interactivity, and Personalisation

Perceived intelligence, perceived interactivity, and perceived personalisation are included in the conceptual model as positive cognitive evaluations that may enhance learners' enjoyment in AI-mediated informal digital learning of English (AI-IDLE). Perceived intelligence refers to learners' beliefs that AI tools can understand their inputs, generate appropriate responses, provide useful feedback, and support English learning in a competent manner (Balakrishnan et al., 2022; Liu, Zou, et al., 2025). When learners perceive AI tools as intelligent, they may experience AI-IDLE as more useful, responsive, and supportive, which can increase their enjoyment during informal English learning (Zadorozhnyy & Lee, 2025).

Perceived interactivity refers to learners' evaluation of the extent to which AI tools enable responsive, reciprocal, and meaningful interaction (F. Wang et al., 2025). Compared with static digital learning resources, AI tools can provide conversational exchanges, answer follow-up questions, and offer immediate responses to learners' language needs (Kim et al., 2022). Such interaction may make AI-IDLE feel more engaging and socially responsive, thereby strengthening learners' enjoyment (AL-Sayid & Kirkil, 2023).

Perceived personalisation refers to learners' beliefs that AI tools can provide content, feedback, explanations, and practice opportunities tailored to their proficiency level, learning goals, interests, and immediate needs (Zhou & Zhang, 2025). In informal and self-regulated learning contexts, personalised support may help learners perceive AI-IDLE as more relevant and adaptive (Liu & Ma, 2024). When learners feel that AI tools address their individual learning needs, they are more likely to experience enjoyment in the learning process (Liu et al., 2024). Accordingly, the following hypotheses are proposed:

H3: Perceived intelligence is positively associated with enjoyment in AI-IDLE.
H4: Perceived interactivity is positively associated with enjoyment in AI-IDLE.
H5: Perceived personalisation is positively associated with enjoyment in AI-IDLE.

### 3.4 Perceived Ineffectiveness, Uncontrollability, and Complexity

Perceived ineffectiveness, perceived uncontrollability, and perceived complexity are included in the conceptual model as negative cognitive evaluations that may increase learners' boredom in AI-mediated informal digital learning of English (AI-IDLE). Perceived ineffectiveness refers to learners' beliefs that AI tools do not adequately support their English learning goals, provide useful feedback, or contribute to meaningful learning progress (Ye et al., 2023). When learners perceive AI-IDLE as ineffective, they may find the learning process less valuable and less stimulating, which can increase boredom (R. Wu, 2023).

Perceived uncontrollability refers to learners' perception that they have limited control over AI-mediated learning processes, outputs, or interactional direction (Hartanto & Thorwart, 2022). In AI-IDLE, uncontrollability may occur when AI tools generate unexpected, irrelevant, inaccurate, or difficult-to-modify responses (Soyoof et al., 2023). When learners feel unable to guide the interaction or obtain the support they need, AI-IDLE may become frustrating, passive, or disengaging, thereby increasing boredom (Liu et al., 2024).

Perceived complexity refers to learners' evaluation that AI tools are difficult to understand, operate, prompt, or integrate into their informal English learning routines (Hmoud et al., 2023). Although AI tools may offer flexible learning opportunities, excessive complexity can impose cognitive burden and reduce learners' willingness to engage deeply with the technology (Capponi et al., 2024). When AI-IDLE requires too much effort to manage, learners may experience reduced interest and increased boredom (Soyoof et al., 2023). Accordingly, the following hypotheses are proposed:

H6: Perceived ineffectiveness is positively associated with boredom in AI-IDLE.

H7: Perceived uncontrollability is positively associated with boredom in AI-IDLE.
H8: Perceived complexity is positively associated with boredom in AI-IDLE.

## 4. Methodology

### 4.1 Research Design

This study adopted a quantitative cross-sectional design to examine learners' intermittent discontinuance in AI-mediated informal digital English learning (AI-IDLE). Guided by the cognition–affect–conation framework, it investigated how cognitive evaluations of AI tools influenced enjoyment and boredom, which in turn predicted intermittent discontinuance. Survey data from Chinese university students with prior AI-IDLE experience were analysed using structural equation modelling (SEM) to test hypothesised relationships and fuzzy-set qualitative comparative analysis (fsQCA) to identify configurational pathways leading to discontinuance.

### 4.2 Participants

Participants were recruited via Credamo, a professional online survey platform in China. To be eligible for inclusion, respondents had to be university students who had prior experience using AI tools for informal English learning beyond formal classroom requirements. In total, 689 questionnaires were submitted. Following data screening, 57 responses were removed because respondents did not meet the inclusion criteria, failed attention-check items, completed the questionnaire in an unrealistically short time, provided patterned responses, or submitted incomplete questionnaires (Ward & Meade, 2023). This yielded a final valid sample of 632 participants. As shown in Table 1, the sample included 287 males ($n$ = 287, 45.4%) and 345 females ($n$ = 345, 54.6%). Regarding age, 218 participants were aged 18–20 ($n$ = 218, 34.5%), 301 were aged 21–23 ($n$ = 301, 47.6%), and 113 were aged 24 or above ($n$ = 113, 17.9%). Most participants were undergraduates ($n$ = 468, 74.1%), while 164 were postgraduates ($n$ = 164, 25.9%). In terms of academic discipline, 296 participants were from STEM fields ($n$ = 296, 46.8%), whereas 336 were from non-STEM fields ($n$ = 336, 53.2%). Before completing the survey, all participants provided informed consent and were informed that participation was voluntary, anonymous, and confidential.

**Table 1. Participant Characteristics ($N$ = 632)**

| Characteristic | Category | *n* | % |
|---|---|---|---|
| Gender | Male | 287 | 45.4 |
| | Female | 345 | 54.6 |
| Age | 18–20 | 218 | 34.5 |
| | 21–23 | 301 | 47.6 |
| | 24 or above | 113 | 17.9 |
| Study level | Undergraduate | 468 | 74.1 |
| | Postgraduate | 164 | 25.9 |
| Academic discipline | STEM | 296 | 46.8 |
| | Non-STEM | 336 | 53.2 |

### 4.3 Measurement

All constructs, perceived intelligence (Balakrishnan et al., 2022), perceived interactivity (F. Wang et al., 2025), perceived personalisation (Zhou & Zhang, 2025), perceived ineffectiveness (Ye et al., 2023), perceived uncontrollability (Hartanto & Thorwart, 2022), perceived complexity (Hmoud et al., 2023), enjoyment (Liu, Zou, et al., 2025), boredom (Li et al., 2023), and AI-IDLE intermittent discontinuance (Zhou & Zhang, 2025), were measured using multi-item reflective scales adapted from prior studies and contextualised for AI-mediated informal digital learning of English (AI-IDLE) (Y. Du, 2024) (see Table 2). All items were rated on a five-point Likert scale ranging from 1 = strongly disagree to 5 = strongly agree. The English items were translated into Chinese and checked through back-translation to ensure semantic equivalence (Klotz et al., 2023). A pilot test was conducted with 30 university students who had experience using AI tools for informal English learning, and minor wording refinements were made based on their feedback. Common method variance (CMV) (Podsakoff et al., 2024) was assessed using Harman's single-factor test. The first unrotated factor explained 31.42% of the total variance, which was below the recommended 50% threshold, suggesting that CMV was unlikely to be a serious concern in this study.

**Table 2. Constructs and Measurement Items**

| Construct | Item | Measurement Item (English) | Measurement Item (Chinese) |
|---|---|---|---|
| Perceived intelligence | PI1 | AI tools can understand my English learning needs. | AI 工具能够理解我的英语学习需求。 |
| | PI2 | AI tools can provide appropriate responses to my English learning questions. | AI 工具能够对我的英语学习问题提供恰当的回应。 |
| | PI3 | AI tools can give intelligent support for my informal English learning. | AI 工具能够为我的非正式英语学习提供智能化支持。 |
| Perceived interactivity | PINT1 | AI tools respond promptly when I interact with them. | 当我与 AI 工具互动时，它们能够及时回应。 |
| | PINT2 | AI tools allow me to have reciprocal interaction during English learning. | AI 工具能够让我在英语学习过程中进行双向互动。 |
| | PINT3 | AI tools enable meaningful interaction for informal English learning. | AI 工具能够为非正式英语学习提供有意义的互动。 |
| Perceived personalisation | PP1 | AI tools provide English learning support tailored to my needs. | AI 工具能够根据我的需求提供英语学习支持。 |
| | PP2 | AI tools adapt their responses to my English proficiency level. | AI 工具能够根据我的英语水平调整其回应。 |
| | PP3 | AI tools provide learning suggestions that fit my personal English learning goals. | AI 工具能够提供符合我个人英语学习目标的学习建议。 |
| Perceived ineffectiveness | PINE1 | AI tools do not effectively support my informal English learning. | AI 工具不能有效支持我的非正式英语学习。 |
| | PINE2 | AI tools do not help me improve my English learning. | AI 工具不能帮助我提升英语学习效果。 |
| | PINE3 | Using AI tools is not useful for my English learning. | 使用 AI 工具对我的英语学习没有帮助。 |
| Perceived uncontrollability | PU1 | I feel that I cannot control the learning process when using AI tools. | 使用 AI 工具时，我觉得自己无法控制学习过程。 |
| | PU2 | I find it difficult to guide AI tools to respond as I expect. | 我发现很难引导 AI 工具按照我的期望进行回应。 |
| | PU3 | AI tools often produce responses that I cannot easily adjust. | AI 工具经常生成我难以调整的回应。 |
| Perceived complexity | PC1 | Learning to use AI tools for informal English learning is difficult for me. | 学会使用 AI 工具进行非正式英语学习对我来说是困难的。 |
| | PC2 | Using AI tools for English learning requires too much effort. | 使用 AI 工具进行英语学习需要投入过多精力。 |
| | PC3 | It is complicated for me to use AI tools effectively for English learning. | 对我来说，有效使用 AI 工具进行英语学习是复杂的。 |
| Enjoyment | ENJ1 | I enjoy using AI tools for informal English learning. | 我喜欢使用 AI 工具进行非正式英语学习。 |
| | ENJ2 | Using AI tools for English learning is pleasant. | 使用 AI 工具进行英语学习是令人愉快的。 |
| | ENJ3 | I feel interested when using AI tools for informal English learning. | 使用 AI 工具进行非正式英语学习时，我感到有兴趣。 |
| Boredom | BOR1 | I feel bored when using AI tools for informal English learning. | 使用 AI 工具进行非正式英语学习时，我感到无聊。 |
| | BOR2 | Using AI tools for English learning feels monotonous. | 使用 AI 工具进行英语学习让我觉得单调。 |
| | BOR3 | I lose interest when using AI tools for informal English learning. | 使用 AI 工具进行非正式英语学习时，我会失去兴趣。 |
| AI-IDLE intermittent discontinuance | AID1 | I sometimes temporarily stop using AI tools for informal English learning, even though I may use them again later. | 我有时会暂时停止使用 AI 工具进行非正式英语学习，尽管之后可能会再次使用。 |
| | AID2 | My use of AI tools for informal English learning is sometimes interrupted for a period of time and then resumed. | 我使用 AI 工具进行非正式英语学习的过程有时会中断一段时间，然后再恢复。 |
| | AID3 | I sometimes drift away from using AI tools for informal English learning, but later return to them. | 我有时会逐渐减少或停止使用 AI 工具进行非正式英语学习，但之后又会重新使用。 |

### 4.4 Data Analysis

Data analysis was conducted in two main stages. First, structural equation modelling (SEM) (Kline, 2023) was used to examine the measurement model and structural model. Descriptive statistics were calculated to assess the distribution of the constructs, and confirmatory factor analysis was performed to evaluate reliability, convergent validity, and discriminant validity. Model fit was assessed using $\chi^2/df$, comparative fit index (CFI), Tucker–Lewis index (TLI), root mean square error of approximation (RMSEA), and standardised root mean square residual (SRMR). The hypothesised paths and mediation effects were then tested using SEM with bootstrapping. Second, fuzzy-set qualitative comparative analysis (fsQCA) (Schneider & Wagemann, 2012) was conducted to identify configurations of cognitive and affective conditions leading to AI-IDLE intermittent discontinuance. The variables were calibrated into fuzzy sets, followed by necessity analysis and truth-table analysis to examine necessary and sufficient conditions for the outcome.

## 5. Results

### 5.1 Structural Equation Modelling (SEM)

Descriptive statistics for the main constructs are presented in Table 3. The mean scores ranged from 2.38 to 3.88, indicating moderate levels of learners' perceptions, affective experiences, and AI-mediated informal digital learning of English intermittent discontinuance (AI-IDLE intermittent discontinuance; AID). The skewness values ranged from -0.55 to 0.52, and the kurtosis values ranged from 0.08 to 0.41, suggesting that the data did not substantially deviate from normality. As shown in Table 4, both the measurement model and the structural model demonstrated acceptable model fit. For the measurement model, $\chi^2/df = 2.41$, comparative fit index (CFI) = 0.932, Tucker–Lewis index (TLI) = 0.921, root mean square error of approximation (RMSEA) = 0.047, and standardised root mean square residual (SRMR) = 0.056. For the structural model, $\chi^2/df = 2.63$, CFI = 0.921, TLI = 0.910, RMSEA = 0.051, and SRMR = 0.061. These values met commonly recommended thresholds, indicating that the proposed model fitted the data adequately.

**Table 3. Descriptive Statistics of the Constructs**

| Construct | *M* | *SD* | Skewness | Kurtosis |
|---|---|---|---|---|
| Perceived intelligence | 3.82 | 0.71 | -0.48 | 0.32 |
| Perceived interactivity | 3.76 | 0.74 | -0.41 | 0.18 |
| Perceived personalisation | 3.69 | 0.77 | -0.36 | 0.11 |
| Perceived ineffectiveness | 2.38 | 0.82 | 0.52 | 0.27 |
| Perceived uncontrollability | 2.56 | 0.79 | 0.39 | 0.14 |
| Perceived complexity | 2.47 | 0.81 | 0.46 | 0.21 |
| Enjoyment | 3.88 | 0.69 | -0.55 | 0.41 |
| Boredom | 2.41 | 0.84 | 0.49 | 0.24 |
| AI-IDLE intermittent discontinuance | 2.73 | 0.86 | 0.31 | 0.08 |

**Table 4. Model Fit Indices**

| Fit Index | Threshold | Measurement Model | Structural Model |
|---|---|---|---|
| $\chi^2$/df | < 3.00 | 2.41 | 2.63 |
| CFI | > 0.90 | 0.932 | 0.921 |
| TLI | > 0.90 | 0.921 | 0.910 |
| RMSEA | < 0.08 | 0.047 | 0.051 |
| SRMR | < 0.08 | 0.056 | 0.061 |

The reliability and convergent validity results are reported in Table 5. All standardised factor loadings exceeded 0.70, ranging from 0.74 to 0.89. Cronbach's $\alpha$ values ranged from 0.77 to 0.82, and composite reliability (CR) values ranged from 0.84 to 0.88, exceeding the recommended threshold of 0.70. The average variance extracted (AVE) values ranged from 0.64 to 0.71, exceeding the threshold of 0.50. These results indicate satisfactory internal consistency reliability and convergent validity. Discriminant validity was further assessed using the Fornell–Larcker criterion and the heterotrait–monotrait ratio (HTMT). As shown in Table 6, the square root of AVE for each construct was greater than its correlations with other constructs. In addition, Table 7 shows that all HTMT values were below 0.85, supporting acceptable discriminant validity.

**Table 5. Reliability and Convergent Validity**

| Construct | Item | Loading | Cronbach's α | CR | AVE |
|---|---|---|---|---|---|
| Perceived intelligence | PI1 | 0.78 | 0.81 | 0.88 | 0.70 |
| | PI2 | 0.85 | | | |
| | PI3 | 0.88 | | | |
| Perceived interactivity | PINT1 | 0.77 | 0.80 | 0.87 | 0.69 |
| | PINT2 | 0.84 | | | |
| | PINT3 | 0.87 | | | |
| Perceived personalisation | PP1 | 0.76 | 0.79 | 0.86 | 0.67 |
| | PP2 | 0.82 | | | |
| | PP3 | 0.87 | | | |
| Perceived ineffectiveness | PINE1 | 0.75 | 0.78 | 0.85 | 0.66 |
| | PINE2 | 0.81 | | | |
| | PINE3 | 0.87 | | | |
| Perceived uncontrollability | PU1 | 0.74 | 0.77 | 0.84 | 0.64 |
| | PU2 | 0.80 | | | |
| | PU3 | 0.86 | | | |
| Perceived complexity | PC1 | 0.76 | 0.79 | 0.86 | 0.67 |
| | PC2 | 0.82 | | | |
| | PC3 | 0.87 | | | |
| Enjoyment | ENJ1 | 0.79 | 0.82 | 0.88 | 0.71 |
| | ENJ2 | 0.85 | | | |
| | ENJ3 | 0.89 | | | |
| Boredom | BOR1 | 0.77 | 0.80 | 0.87 | 0.69 |
| | BOR2 | 0.84 | | | |
| | BOR3 | 0.87 | | | |
| AI-IDLE intermittent discontinuance | AID1 | 0.78 | 0.81 | 0.88 | 0.70 |

| | |
|---|---|
| AID2 | 0.84 |
| AID3 | 0.89 |

**Table 6. Discriminant Validity (Fornell–Larcker Criterion)**

| Construct | PI | PINT | PP | PINE | PU | PC | ENJ | BOR | AID |
|---|---|---|---|---|---|---|---|---|---|
| PI | **0.84** | | | | | | | | |
| PINT | 0.56 | **0.83** | | | | | | | |
| PP | 0.53 | 0.58 | **0.82** | | | | | | |
| PINE | -0.32 | -0.35 | -0.34 | **0.81** | | | | | |
| PU | -0.29 | -0.31 | -0.33 | 0.55 | **0.80** | | | | |
| PC | -0.30 | -0.34 | -0.32 | 0.57 | 0.59 | **0.82** | | | |
| ENJ | 0.52 | 0.55 | 0.57 | -0.38 | -0.34 | -0.36 | **0.84** | | |
| BOR | -0.33 | -0.36 | -0.35 | 0.58 | 0.56 | 0.60 | -0.42 | **0.83** | |
| AID | -0.28 | -0.30 | -0.31 | 0.46 | 0.44 | 0.47 | -0.39 | 0.51 | **0.84** |

Note. Diagonal values in bold represent the square roots of AVE, and off-diagonal values represent inter-construct correlations.

**Table 7. Discriminant Validity (Heterotrait–Monotrait Ratio)**

| Construct | PI | PINT | PP | PINE | PU | PC | ENJ | BOR | AID |
|---|---|---|---|---|---|---|---|---|---|
| PI | — | | | | | | | | |
| PINT | 0.68 | — | | | | | | | |
| PP | 0.65 | 0.70 | — | | | | | | |
| PINE | 0.39 | 0.42 | 0.41 | — | | | | | |
| PU | 0.36 | 0.38 | 0.40 | 0.67 | — | | | | |
| PC | 0.37 | 0.41 | 0.39 | 0.69 | 0.71 | — | | | |
| ENJ | 0.63 | 0.66 | 0.68 | 0.46 | 0.41 | 0.43 | — | | |
| BOR | 0.40 | 0.43 | 0.42 | 0.70 | 0.68 | 0.72 | 0.50 | — | |
| AID | 0.34 | 0.36 | 0.37 | 0.55 | 0.53 | 0.56 | 0.47 | 0.61 | — |

**Figure 2. Structural Model Results**

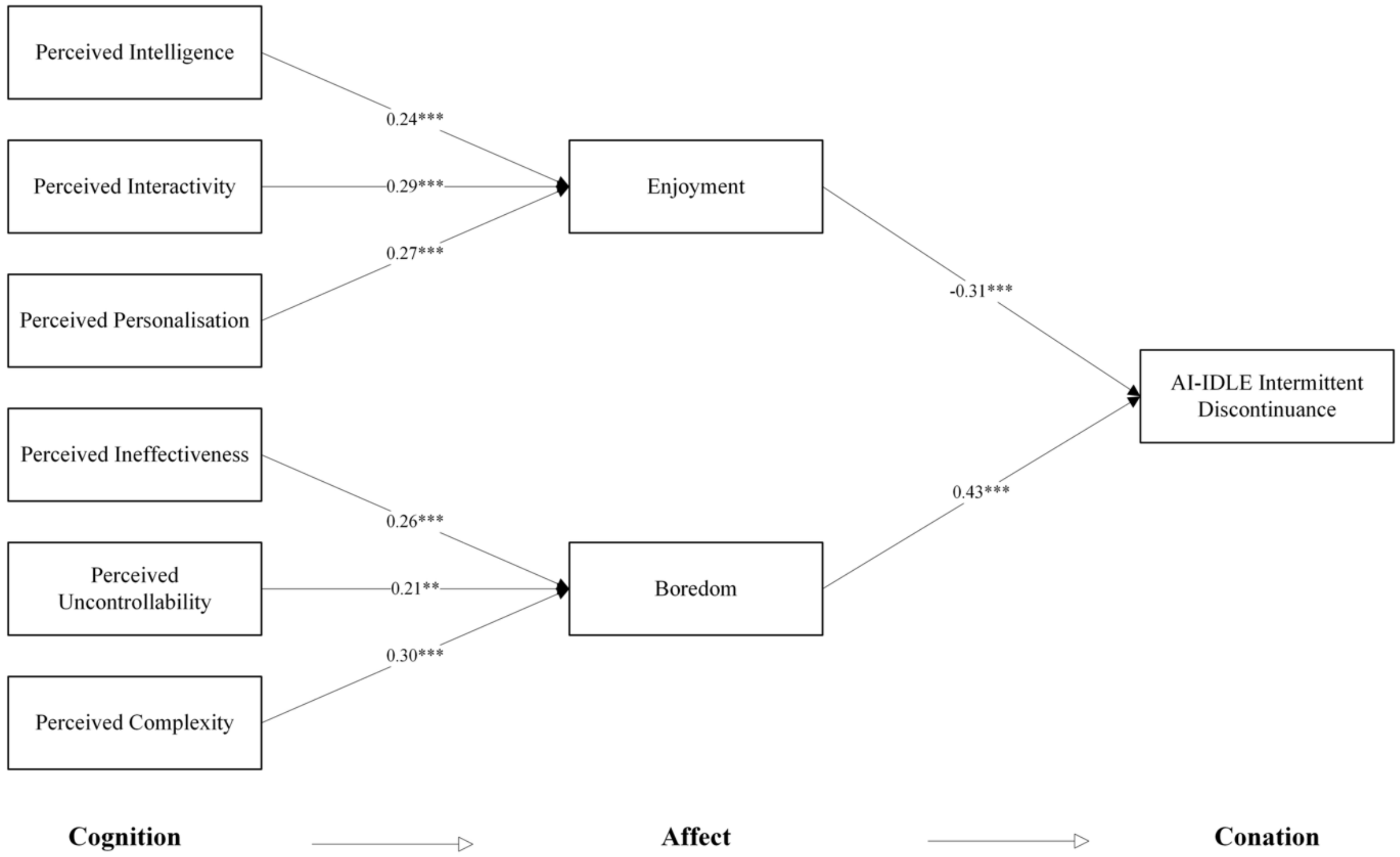

Note. *** $p < 0.001$; ** $p < 0.01$.

The structural model results are presented in Table 8. Enjoyment (ENJ) was negatively associated with AID ($\beta$ = -0.31, *SE* = 0.052, *z* = -5.96), supporting H1. Boredom (BOR) was positively associated with AID ($\beta$ = 0.43, *SE* = 0.049, *z* = 8.78), supporting H2. Regarding the antecedents of ENJ, perceived intelligence (PI; $\beta$ = 0.24, *SE* = 0.046, *z* = 5.22), perceived interactivity (PINT; $\beta$ = 0.29, *SE* = 0.048, *z* = 6.04), and perceived personalisation (PP; $\beta$ = 0.27, *SE* = 0.047, *z* = 5.74) were all positively associated with ENJ, supporting H3, H4, and H5. Regarding the antecedents of BOR, perceived ineffectiveness (PINE; $\beta$ = 0.26, *SE* = 0.050, *z* = 5.20), perceived uncontrollability (PU; $\beta$ = 0.21, *SE* = 0.067, *z* = 3.13), and perceived complexity (PC; $\beta$ = 0.30, *SE* = 0.055, *z* = 5.45) were positively associated with BOR, supporting H6, H7, and H8.

**Table 8. Structural Model Results**

| Hypothesis | Path | $\beta$ | *SE* | *z* | Result |
|---|---|---|---|---|---|
| H1 | ENJ → AID | -0.31*** | 0.052 | -5.96 | Supported |
| H2 | BOR → AID | 0.43*** | 0.049 | 8.78 | Supported |
| H3 | PI → ENJ | 0.24*** | 0.046 | 5.22 | Supported |
| H4 | PINT → ENJ | 0.29*** | 0.048 | 6.04 | Supported |
| H5 | PP → ENJ | 0.27*** | 0.047 | 5.74 | Supported |
| H6 | PINE → BOR | 0.26*** | 0.050 | 5.20 | Supported |
| H7 | PU → BOR | 0.21** | 0.067 | 3.13 | Supported |
| H8 | PC → BOR | 0.30*** | 0.055 | 5.45 | Supported |

Note. *** $p < 0.001$; ** $p < 0.01$.

The mediation analysis was conducted using bootstrapping with 5,000 resamples, and the results are reported in Table 9. The indirect effects of PI, PINT, and PP on AID through ENJ were significant because their 95% bootstrap confidence intervals (CIs) did not include zero. Specifically, the indirect effects were PI → ENJ → AID = -0.074, 95% CI [-0.112, -0.042]; PINT → ENJ → AID = -0.090, 95% CI [-0.134, -0.052]; and PP → ENJ → AID = -0.084, 95% CI [-0.127, -0.048]. Similarly, the indirect effects of PINE, PU, and PC on AID through BOR were significant: PINE → BOR → AID = 0.112, 95% CI [0.066, 0.168]; PU → BOR → AID = 0.090, 95% CI [0.035, 0.153]; and PC → BOR → AID = 0.129, 95% CI [0.076, 0.190]. Overall, the SEM results support the proposed cognition–affect–conation model, indicating that positive cognitive evaluations reduce AID indirectly through enjoyment, whereas negative cognitive evaluations increase AID indirectly through boredom.

**Table 9. Mediation Analysis Results (Bootstrapping)**

| Indirect path | Indirect effect | *SE* | 95% CI | Result |
|---|---|---|---|---|
| PI → ENJ → AID | -0.074 | 0.018 | [-0.112, -0.042] | Supported |
| PINT → ENJ → AID | -0.090 | 0.021 | [-0.134, -0.052] | Supported |
| PP → ENJ → AID | -0.084 | 0.020 | [-0.127, -0.048] | Supported |
| PINE → BOR → AID | 0.112 | 0.026 | [0.066, 0.168] | Supported |
| PU → BOR → AID | 0.090 | 0.030 | [0.035, 0.153] | Supported |
| PC → BOR → AID | 0.129 | 0.029 | [0.076, 0.190] | Supported |

### 5.2 Fuzzy-Set Qualitative Comparative Analysis (fsQCA)

In addition to structural equation modelling (SEM), fuzzy-set qualitative comparative analysis (fsQCA) was conducted to examine the configurational mechanisms underlying AI-IDLE intermittent discontinuance. Before the analysis, all antecedent conditions and the outcome were calibrated into fuzzy-set membership scores using three qualitative anchors: the 95th percentile for full membership, the 50th percentile as the crossover point, and the 5th percentile for full non-membership. The antecedent conditions included perceived intelligence (PI), perceived interactivity (PINT), perceived personalisation (PP), perceived ineffectiveness (PINE), perceived uncontrollability (PU), perceived complexity (PC), enjoyment (ENJ), and boredom (BOR), while the outcome was AI-IDLE intermittent discontinuance (AID). A necessity analysis was first performed to determine whether any single condition was necessary for AID. The results showed that no individual condition reached the commonly recommended consistency threshold of 0.90, indicating that AID was unlikely to be explained by any single necessary antecedent condition. After the necessity analysis, a truth table was constructed, and consistency and frequency thresholds were applied to identify sufficient configurations leading to AID.

As shown in Table 10, four configuration paths were identified, with an overall solution consistency of 0.84 and an overall solution coverage of 0.68. Path 1 shows that the core presence of PINE and BOR, together with the core absence of PI and PINT, can lead to AID, suggesting that learners may temporarily withdraw from AI-IDLE when they perceive AI tools as ineffective and boring while lacking positive perceptions of intelligence and interactivity. Path 2 indicates that the core presence of PINE, PC, and BOR, combined with the core absence of PP, constitutes another sufficient path to AID. Path 3 highlights a control–complexity mechanism, in which the core presence of PU and PC, together with the core absence of ENJ, leads to AID. Path 4 further shows that the core presence of PU and BOR is sufficient for AID when accompanied by the peripheral presence of PC and the peripheral absence of PP. Overall, these findings complement the SEM results by demonstrating equifinality: AI-IDLE intermittent discontinuance can emerge from multiple combinations of cognitive and affective conditions rather than from a single isolated factor.

**Table 10. Configuration Paths of AI-IDLE Intermittent Discontinuance**

| Condition | Path 1 | Path 2 | Path 3 | Path 4 |
|---|---|---|---|---|
| Perceived intelligence | ⊗ | | ⊖ | |
| Perceived interactivity | ⊗ | ⊖ | | |
| Perceived personalisation | ⊖ | ⊗ | | ⊖ |
| Perceived ineffectiveness | ● | ● | ○ | |
| Perceived uncontrollability | ○ | | ● | ● |
| Perceived complexity | | ● | ● | ○ |
| Enjoyment | ⊗ | ⊖ | ⊗ | |
| Boredom | ● | ● | ○ | ● |
| Raw consistency | 0.86 | 0.84 | 0.82 | 0.81 |
| PRI consistency | 0.79 | 0.77 | 0.75 | 0.73 |
| Raw coverage | 0.34 | 0.29 | 0.24 | 0.21 |
| Unique coverage | 0.12 | 0.09 | 0.07 | 0.06 |
| Overall solution consistency | 0.84 | | | |
| Overall solution coverage | 0.68 | | | |

Note. ● indicates the presence of a core condition; ○ indicates the presence of a peripheral condition; ⊗ indicates the absence of a core condition; ⊖ indicates the absence of a peripheral condition; blank cells indicate “don’t care” conditions. PRI = proportional reduction in inconsistency.

## 6. Discussion

### 6.1 Net Effects of Cognitive and Affective Factors on AI-IDLE Intermittent Discontinuance

The SEM results provide empirical support for the cognition–affect–conation framework, indicating that learners’ cognitive evaluations of AI tools shape affective experiences, which subsequently influence AI-IDLE intermittent discontinuance. Specifically, enjoyment was negatively associated with intermittent discontinuance, whereas boredom was positively associated with intermittent discontinuance. This suggests that learners are less likely to temporarily withdraw from AI-mediated informal digital English learning when they experience AI tools as pleasant, interesting, and intrinsically rewarding. Conversely, when AI-IDLE becomes monotonous, unstimulating, or disengaging, learners are more likely to suspend use temporarily. These findings align with prior research showing that enjoyment promotes sustained informal digital English learning engagement (Lee et al., 2024; Liu, Zou, et al., 2025), while boredom weakens learners’ willingness to maintain language-learning activities (Li et al., 2023; Zhang & Li, 2025).

The results further show that positive cognitive evaluations reduced intermittent discontinuance indirectly through enjoyment. Perceived intelligence, perceived interactivity, and perceived personalisation were all positively associated with enjoyment, which in turn negatively predicted AI-IDLE intermittent discontinuance. Among these antecedents, perceived interactivity showed the strongest effect on enjoyment, followed by perceived personalisation and perceived intelligence. This indicates that learners derive enjoyment not only from the functional competence of AI tools, but also from their capacity to provide reciprocal interaction and tailored support. These findings are consistent with previous studies suggesting that intelligent, interactive, and personalised AI systems can enhance learners' engagement and positive emotional experience in AI-IDLE (Balakrishnan et al., 2022; Wang et al., 2025; Zhou & Zhang, 2025; Zadorozhnyy & Lee, 2025). Therefore, enjoyment appears to function as a protective affective mechanism through which favourable perceptions of AI tools reduce learners' tendency to drift away from AI-IDLE.

Negative cognitive evaluations, by contrast, increased intermittent discontinuance indirectly through boredom. Perceived ineffectiveness, perceived uncontrollability, and perceived complexity were all positively associated with boredom, which subsequently increased AI-IDLE intermittent discontinuance. Perceived complexity had the strongest effect on boredom, followed by perceived ineffectiveness and perceived uncontrollability, suggesting that learners are especially likely to become bored when AI-IDLE requires excessive effort, is difficult to manage, or fails to produce meaningful learning value. This finding extends prior work on perceived learning ineffectiveness, uncontrollability, and technological complexity (Hartanto & Thorwart, 2022; Hmoud et al., 2023; Ye et al., 2023) by showing that these negative evaluations operate through affective disengagement. Overall, the net-effect results indicate that AI-IDLE intermittent discontinuance is not merely a behavioural response to technological limitations, but an affectively mediated outcome: positive perceptions reduce discontinuance by enhancing enjoyment, whereas negative perceptions increase discontinuance by intensifying boredom.

### 6.2 Configurational Mechanisms Underlying AI-IDLE Intermittent Discontinuance

The fsQCA results complement the SEM findings by showing that AI-IDLE intermittent discontinuance arises through multiple configurational pathways rather than through any single antecedent condition. The necessity analysis indicated that no individual cognitive or affective factor was necessary for intermittent discontinuance, which supports the principle of equifinality in configurational analysis (Schneider & Wagemann, 2012). Four sufficient configurations were identified, with an overall solution consistency of 0.84 and coverage of 0.68, indicating that different combinations of unfavourable cognitive evaluations and affective disengagement can explain learners' temporary withdrawal from AI-IDLE. In particular, perceived ineffectiveness and boredom appeared as central conditions in Paths 1 and 2, suggesting that learners may discontinue AI-IDLE when they perceive AI tools as offering limited learning value and when their use becomes monotonous or unstimulating. This finding is consistent with prior work showing that perceived learning ineffectiveness can undermine engagement (Ye et al., 2023) and that boredom is a salient negative emotion in language learning that weakens persistence (Li et al., 2023; Zhang & Li, 2025).

The remaining configurations highlight a control–complexity mechanism. Paths 3 and 4 show that perceived uncontrollability and perceived complexity can jointly lead to intermittent discontinuance, especially when enjoyment is absent or boredom is present. This suggests that learners may drift away from AI-IDLE not only because AI tools are perceived as ineffective, but also because they are difficult to guide, adjust, or integrate into informal English learning routines. Such findings extend previous research on AI-IDLE and technology-mediated language learning by demonstrating that discontinuance is shaped by conjunctural causation: different cognitive barriers may combine with affective states to produce the same behavioural outcome (Soyoof et al., 2023; Zhou & Zhang, 2025). Overall, the configurational results indicate that AI-IDLE intermittent discontinuance should be understood as a heterogeneous post-adoption phenomenon. Some learners may withdraw because AI tools feel ineffective and boring, whereas others may disengage because the learning process feels uncontrollable, complex, and insufficiently enjoyable. This nuanced account supports the cognition–affect–conation perspective by showing that cognitive appraisals and affective experiences operate together in shaping discontinuous AI-IDLE engagement (Zhou & Zhang, 2024).

### 6.3 Theoretical and Practical Implications

This study offers several theoretical implications. First, it extends AI-IDLE research beyond adoption, acceptance, and learning benefits by positioning intermittent discontinuance as a distinct post-adoption outcome. This is important because learners' engagement with AI tools in informal English learning is often voluntary, unstable, and context-dependent. Second, by applying the cognition–affect–conation framework, the study shows that cognitive evaluations do not influence discontinuance only directly; rather, they operate through affective experiences. Positive perceptions of intelligence, interactivity, and personalisation reduce discontinuance by enhancing enjoyment, whereas perceived ineffectiveness, uncontrollability, and complexity increase discontinuance by intensifying boredom. Third, the integration of SEM and fsQCA provides both net-effect and configurational evidence, showing that AI-IDLE intermittent discontinuance is produced through multiple cognitive–affective mechanisms rather than a single linear pathway.

Practically, the findings suggest that designers, educators, and AI platform developers should reduce learners' temporary withdrawal by strengthening enjoyable and controllable AI-IDLE experiences. AI tools should provide accurate feedback, meaningful interaction, adaptive learning support, and transparent response mechanisms so that learners perceive them as intelligent, interactive, personalised, and manageable. At the same time, unnecessary complexity should be reduced through clearer interfaces, prompt guidance, task scaffolding, and learner-friendly explanations. Educators can also help by modelling effective AI use, recommending suitable AI-IDLE activities, and teaching learners how to evaluate, refine, and control AI-generated outputs. These practices may reduce boredom, increase enjoyment, and support more stable engagement with AI-mediated informal English learning.

### 6.4 Limitations and Future Directions

This study has several limitations. First, the cross-sectional design limits causal interpretation, as the data capture learners' cognitive evaluations, affective experiences, and intermittent discontinuance at one point in time (Y. Du, Yuan, et al., 2026; Tang, Jia, et al., 2026; C. Wang et al., 2026; W. Zhang et al., 2026). Future studies could use longitudinal, diary, or experience-sampling designs to examine how learners move between AI-IDLE use, temporary withdrawal, and reuse over time (Y. Du, Li, et al., 2026; Y. Du, Tang, et al., 2026; Y. Du & He, 2026c; Tang, Lau, et al., 2026). Second, the sample consisted of Chinese university EFL learners, which may limit the generalisability of the findings to other age groups, educational levels, cultural contexts, and language-learning environments (Y. Du, 2025b, 2026; Y. Du & He, 2026b, 2026a). Future research should test the model with more diverse learner populations and compare whether the mechanisms of intermittent discontinuance differ across contexts (C. Du et al., 2025; Y. Du, 2025a; Y. Du et al., 2025; C. Wang et al., 2024).

Third, the study relied on self-reported questionnaire data, which may be affected by recall bias or social desirability bias, even though procedural and statistical checks were applied (Y. Du, 2024; Y. Du et al., 2024; He & Du, 2024; Zou et al., 2024). Future work could combine surveys with learning analytics, AI-use logs, interviews, or screen-recorded interaction data to capture discontinuance more objectively and in greater depth (Chen et al., 2022; Y. Du, 2023; Zou et al., 2023). Finally, this study focused on selected cognitive and affective factors within the cognition–affect–conation framework. Future research could incorporate additional variables, such as self-regulation, trust, AI literacy, privacy concerns, perceived accuracy, learner autonomy, and task type, to develop a more comprehensive explanation of why learners intermittently discontinue AI-mediated informal digital English learning.

## 7. Conclusion

This study examined AI-IDLE intermittent discontinuance through the cognition–affect–conation framework by integrating SEM and fsQCA. The SEM results showed that positive cognitive evaluations, including perceived intelligence, interactivity, and personalisation, reduced intermittent discontinuance indirectly through enjoyment, whereas negative evaluations, including perceived ineffectiveness, uncontrollability, and complexity, increased intermittent discontinuance indirectly through boredom.

The fsQCA results further revealed multiple configurational pathways to discontinuance, indicating that learners may temporarily withdraw from AI-IDLE through different combinations of cognitive barriers and affective disengagement. Overall, the findings suggest that AI-IDLE engagement is dynamic and fragile rather than linear or stable, and that sustaining learners' use of AI tools requires both cognitively supportive and affectively engaging learning experiences.